\documentclass[12pt]{article}
\usepackage{amsmath}
\usepackage{amsfonts}
\usepackage{graphicx}
\usepackage{cite}
\usepackage{bbm}

\newcommand{\nn}{\nonumber}
\newcommand{\Eqn}[1]{&\hspace{-0.5em}#1\hspace{-0.5em}&}
\newcommand{\eqb}{\begin{eqnarray}}
\newcommand{\eqe}{\end{eqnarray}}

\def\ep{\epsilon}

\def\bra{\langle}
\def\bbra{\langle \! \langle}
\def\ket{\rangle}
\def\kket{\rangle \! \rangle}
\def\cdots {\cdot\cdot\cdot}
\def\pint#1 {- \!\!\!\!\!\!\!\! \,\int_{#1}}

\def\comma      { \, , }
\def\period     { \, . }
\def\del        {  \partial  }

\def\half       {\frac{1}{2}}
\def\abs#1      {  \, \vert #1 \vert \,   }
\def\binom#1#2 { \vecii{ {}_{#1} }{\raisebox{.5ex}{$ {}^{#2} $}} }

\def\zbar   {\bar{z}}
\def\wbar   {\bar{w}}

\def\calD   {{\cal D}}

\newcommand{\One}{ {\bf 1}}
\newcommand{\bbC}{{\mathbb C}}
\newcommand{\bbR}{{\mathbb R}}
\newcommand{\bbZ}{{\mathbb Z}}
\newcommand{\bbI}{ {\cal  I}  }

\DeclareMathOperator{\Li}{Li}

\DeclareMathOperator{\tr}{tr}

\DeclareMathOperator{\diag}{diag}

\def\vecii#1#2     {{ #1 \choose #2 }  }
\def\veciii#1#2#3  {\left(\begin{array}{c}#1\\#2\\#3\end{array}\right)}
\def\matrixii#1#2#3#4            {\Bigl( \begin{array}{cc}#1&#2\\#3&#4
                                   \end{array} \Bigr) }
\def\matrixiii#1#2#3#4#5#6#7#8#9 {\left(\begin{array}{ccc}#1&#2&#3\\
                                #4&#5&#6\\#7&#8&#9\end{array}\right)}
\renewcommand{\thesection}
{\arabic{section} \hspace{-.5em}
}
\renewcommand{\thesubsection}
{\arabic{section}.\arabic{subsection}   \hspace{-.5em}
}
\renewcommand{\thesubsubsection}
{\arabic{section}.\arabic{subsection}.\arabic{subsubsection} \hspace {-.5em}
 }

\makeatletter
\renewcommand\section{
\@startsection{section}{3}{\z@}%
{-4.5ex\@plus -1ex \@minus -.2ex}%
{1.5ex \@plus .2ex}%
{\normalfont\large\bfseries\mathversion{bold}}}
\renewcommand\subsection{
\@startsection{subsection}{3}{\z@}%
{-3ex\@plus -1ex \@minus -.2ex}%
{0.7ex \@plus .2ex}%
{\normalfont\normalsize\bfseries\mathversion{bold}}}
\renewcommand\subsubsection{
\@startsection{subsubsection}{3}{\z@}%
{-3.25ex\@plus -1ex \@minus -.2ex}%
{1.5ex \@plus .2ex}%
{\normalfont\normalsize\itshape}}
\makeatother

\makeatletter \@addtoreset{equation}{section} \makeatother
\renewcommand{\theequation}{\arabic{section}.\arabic{equation}}

\makeatletter
\renewcommand{\appendix}{
\renewcommand{\thesection}{\Alph{section}  \hspace{-.5em}}
\renewcommand{\thesubsection}
{\Alph{section}.\arabic{subsection} \hspace{-.5em}}
\renewcommand{\thesubsubsection}
{\Alph{section}.\arabic{subsection}.\arabic{subsubsection} \hspace  {-.5em}}
\@addtoreset{equation}{subsection}
\renewcommand{\theequation}{\Alph{section}.\arabic{equation}}
\setcounter{section}{0}}
\makeatother

\renewcommand{\thefootnote}{\fnsymbol{footnote}}

\def\sigmabar   {\bar{\sigma}}

\def\thetabar   {\bar{\theta}}

\def\adot  {\dot{a}}
\def\bdot  {\dot{b}}
\def\dotc  {\dot{c}}
\def\ddot  {\dot{d}}
\def\alphatil {\tilde{\alpha}}
\def\tilalpha {\tilde{\alpha}}
\def\betatil {\tilde{\beta}}
\def\Stil {\tilde{S}}
\def\Qtil {\tilde{Q}}

\def\Xtil {\tilde{X}}

\def\ktil {\tilde{k}}
\def\Ptil {\tilde{P}}

\def\Ltil {\tilde{L}}
\def\Ttil {\tilde{T}}

\def\calB   {{\cal B}}
\def\calN   {{C}}
\def\S   {{\cal S}}

\begin{document}
\def\papertitlepage{\baselineskip 3.5ex \thispagestyle{empty}}
\def\preprinumber#1#2#3{\hfill \begin{minipage}{2.6cm} #1
                \par\noindent #2
              \par\noindent #3
             \end{minipage}}
\renewcommand{\thefootnote}{\fnsymbol{footnote}}
\newcounter{aff}
\renewcommand{\theaff}{\fnsymbol{aff}}
\newcommand{\affiliation}[1]{
\setcounter{aff}{#1} $\rule{0em}{1.2ex}^\theaff\hspace{-.4em}$}
%
%
\papertitlepage
\setcounter{page}{0}
\preprinumber{}{UTHEP-639}{}
\baselineskip 0.8cm
\vspace*{2.5cm}
\begin{center}
{\large\bf On supersymmetric interfaces for string theory}
\end{center}
\vskip 4ex
\baselineskip 1.0cm
\begin{center}
       
         Yuji  ~Satoh\footnote[3]{\tt ysatoh@het.ph.tsukuba.ac.jp}

\vskip 2ex
 
    {\it Institute of Physics, University of Tsukuba} \\
    \vskip -1.5ex
    {\it Tsukuba, Ibaraki 305-8571, Japan}
\end{center}
\vskip 16ex
%
\baselineskip=3.5ex

\begin{center} {\bf Abstract} \end{center}

\par\medskip
\ 
We construct the world-sheet interface which preserves space-time
supersymmetry in type II  superstring theories in the Green-Schwarz
formalism.
This is an analog of the conformal interface in two-dimensional
conformal field theory. We show that a class of the supersymmetric
interfaces generates T-dualities of type II theories, and that
these interfaces have a geometrical interpretation in the doubled
target space. We compute the partition function
with a pair of the supersymmetric interfaces inserted, from which we read off
the spectrum of the modes coupled to the interfaces
and the Casimir energy between them.
We also derive the transformation rules  
under which a set of D-branes is transformed to another by the interface.

%
%
%
%
%

\vspace*{\fill}
\noindent
December 2011

\newpage
\renewcommand{\thefootnote}{\arabic{footnote}}
\setcounter{footnote}{0}
\setcounter{section}{0}
\baselineskip = 3.3ex
\pagestyle{plain}
\section{Introduction}

Since its discovery, the D-brane has been a central subject  
in the study of superstring theory. On one hand, it preserves
space-time supersymmetry and, on the other, it preserves
world-sheet conformal invariance. 
As is generally the case for those  which preserve
fundamental symmetries, the D-brane plays an important and 
fundamental role in the theory. From the world-sheet point of view,
a natural generalization of 
the D-brane or the conformal boundary/boundary state  
is the conformal interface \cite{Oshikawa:1996dj,Petkova:2000ip,Bachas:2001vj}. 
It is a one-dimensional domain wall/defect
in the world-sheet which preserves the conformal invariance and glues
two generally different conformal field theories (CFTs).
As anticipated,  the conformal interface has interesting properties: 
it generates symmetries of CFT including T-dualities \cite{Frohlich:2004ef},
and transforms a set of D-branes to another \cite{Graham:2003nc,Bachas:2004sy}. 
For the aspects of the conformal interface, we refer to 
\cite{Quella:2002ct,Recknagel:2002qq,Bachas,Fuchs:2007fw,
Fuchs:2007tx,Bachas:2007td,Brunner:2008fa,
Gang:2008sz,Sakai:2008tt,Sarkissian:2008dq,Kapustin:2010zc,Suszek:2011hg}
and references therein.

One can thus expect  that, once embedded in superstring theory, 
the interface would provide an important element 
in order to explore non-perturbative aspects and symmetries 
of superstring theory.
The purpose of this paper is to take a step toward this direction.
In particular, we will study  
the world-sheet interface in type II superstring theories 
in flat space-time in the Green-Schwarz (GS) formalism.
The reason to work in the GS formalism is two-fold. First, 
space-time supersymmetry is manifest and
one can avoid complications due to ghosts, as usual.
Second,  a difficulty has been pointed out \cite{Bachas:2001vj} 
for the conformal interface in the world-sheet of strings:
the interface generally may not preserve enough 
Virasoro symmetries to remove negative norm states, 
except in  the special cases where two sets of the Virasoro symmetries 
are preserved.
In the GS formalism, the physical space is manifestly unitary,  and 
hence this formalism should  provide a safe framework in which one can study
the object whose properties are yet to be investigated. 
As the genus of the world-sheet increases, the interface can be wrapped on non-equivalent
one-cycles. In order for the interface to make full sense in string theory,
it is necessary to clarify how to perform the summation over the genus for correlation 
functions involving interfaces.%
\footnote{
The author would like to thank the referee for pointing out this.} 
This is an important issue for future, and we focus on fixed genus
in this paper.

In the GS formulation, 
the conformal boundary state describing the D-brane in the covariant formulation
is represented as the boundary state preserving space-time 
supersymmetry \cite{Green:1996um}. Similarly, 
the conformal interface would be represented in the GS formulation 
as the interface preserving the space-time supersymmetry.
In this paper, we indeed construct the world-sheet interface with this property.
Since the world-sheet theory in the GS formulation is not a conformal
field theory due to the gauge fixing, 
we call the above boundary states/interfaces the supersymmetric
boundary states/interfaces.
We find two classes of the supersymmetric interfaces, which are 
regarded as generalizations of the $c=1$ permeable conformal interfaces 
\cite{Bachas:2001vj}. 
One describes factorized D-branes/boundary states, whereas the other 
is an analog of  the topological conformal interface \cite{Petkova:2000ip}. 
We show that the latter class generates 
T-dualities of type II theories. 
In both cases, two sets of the Virasoro generators in the physical space 
are preserved, and the difficulty mentioned above may be evaded.
We also study properties of the supersymmetric interface.
First, in parallel with the topological conformal interface \cite{Fuchs:2007fw}, 
we see that 
the corresponding supersymmetric interface has a geometrical interpretation 
in the doubled target space. 
Second,  we compute  the coupling of the massless fields through the interface, 
and confirm  the Buscher rules at the linearized level
in the case of the analog of the topological interface.
Third, we compute  the partition function with a pair of an interface 
and its conjugate
inserted, from which we read off the spectrum of the modes coupled to the interfaces
and the Casimir energy between them.
Finally, we derive the transformations of the D-branes
by the interface. Our results confirm that  the conformal interface is 
embedded into superstring theory at least for fixed genus, though 
somewhat in disguise in our formulation.

The rest of this  paper is organized as follows. In section 2, we summarize
the supersymmetric boundary state for type II superstrings in the
GS formalism,  together with the unfolding procedure of boundary states
to interfaces.
In section 3, we construct the supersymmetric interfaces.
In section 4, we study the target-space geometry and the coupling
of the massless fields. In section 5, we compute the partition 
function with the interfaces inserted. In section 6, we derive the
transformations of the D-branes. We conclude with a summary 
and discussion in section 7.

\section{Supersymmetric boundary states}

The conformal interface in two-dimensional conformal field theory 
is obtained from the conformal boundary state by the unfolding 
procedure \cite{Oshikawa:1996dj,Bachas:2001vj}.  
In type II superstring theories in the Green-Schwarz formalism in light-cone gauge, 
the conformal symmetry is fixed, 
and the guiding principle to construct the boundary state, 
the conformal invariance, is replaced by the invariance under the space-time
supersymmetry. Accordingly,
the conformal boundary state, corresponding to the D-brane, 
is realized as the supersymmetric boundary state preserving  space-time
supersymmetry. 
It is  then expected that  the supersymmetric 
interface in the GS formalism is obtained  by unfolding the supersymmetric
boundary state. This is the strategy which we take in the following.
We thus start our discussion with a summary on the supersymmetric
boundary state  \cite{Green:1996um} and the unfolding procedure.

\subsection{Supersymmetric boundary states in type IIB theory}

To be concrete, we first concentrate on  type IIB theory in flat space-time.
In the GS formalism in light-cone gauge,
one of the light-cone string coordinates is parametrized as
$X^{+} = x^{+} + p^{+} \tau$. The physical degrees of freedom 
are given by eight transverse coordinates, $X^{I}(\tau, \sigma)$ $(I = 1, ..., 8)$, 
and two right- and left-moving SO(8) Majorana-Weyl 
spinors with the same chirality, $S^{a}(\tau-\sigma)$ and $\Stil^{a}(\tau+\sigma)$
$(a=1, ..., 8)$. The other light-cone coordinate $X^{-}$ is determined through
the constraints, 
\eqb\label{lcVirasoro}
 p^{+} \del_{\pm} X^{-} = \del_{\pm}X_{I}\del_{\pm}X^{I} 
      +\frac{i}{2} S^{a}\del_{-} S^{a}  + \frac{i}{2} \Stil^{a} \del_{+} \Stil^{a}\period 
\eqe
The half of the space-time supersymmetry is realized linearly  by
 the spinor zero modes, 
\eqb
   Q^{a} := \sqrt{2p^{+}} S_{0}^{a} \comma \quad
   \Qtil^{a} := \sqrt{2p^{+}} \Stil_{0}^{a} \comma
\eqe
whereas the other half is realized non-linearly by 
\eqb\label{NLQ}
   Q^{\adot} := \frac{1}{ \sqrt{p^{+}} } \sigma_{a\adot}^{I} \sum_{n = -\infty}^{\infty}
       S_{-n}^{a} \alpha_{n}^{I}  \comma \quad
   \Qtil^{\adot} := \frac{1}{\sqrt{p^{+}} } \sigma_{a\adot}^{I} \sum_{n = -\infty}^{\infty}
       \Stil_{-n}^{a}\alphatil_{n}^{I}  \period
\eqe
The modes of the fields satisfy the relations, 
\eqb
   [\alpha_{m}^{I}, \alpha_{n}^{J}] = m \delta_{m+n,0} \delta^{IJ}\comma
   \quad 
   \{ S_{m}^{a}, S_{n}^{b}\} = \delta_{m+n,0} \delta^{ab} \comma
\eqe
and similar ones for $\alphatil_{n}^{I}, \Stil_{n}^{a}$. The matrices 
$\sigma^{I}_{a\adot}$ together with $\sigmabar^{I}_{\adot a} (= \sigma^{I}_{a\adot})$
form eight-dimensional gamma matrices 
\eqb
   \gamma^{I} =  \biggl(\begin{array}{cc}0 & \sigma^{I}  \\\sigmabar^{I} & 0\end{array}\biggr)
   \quad 
   {\rm satisfying} \quad 
  \{ \gamma^{I}, \gamma^{J} \} = 2 \delta^{IJ} \period
\eqe  
The anti-commutation relations among the supercharges are, e.g., 
\eqb
  \{ Q^{a}, Q^{b} \} = 2p^{+} \delta^{ab} \comma \quad
  \{ Q^{a}, Q^{\adot} \} =  \sqrt{2}
    \alpha_{0}^{I}\sigma^{I}_{a\adot} \comma \quad
    \{ Q^{\adot}, Q^{\bdot} \} = P^{-}
    \delta^{\adot\bdot} \comma
\eqe
where
\eqb
   P^{-} = \frac{2}{p^{+}} \Bigl( \half  \alpha_{0}^{I} \alpha_{0}^{I} 
   +  N_{\rm b} + N_{\rm f}\Bigr) 
   \comma
\eqe
and
\eqb
 N_{\rm b} = \sum_{n=1}^{\infty} \alpha_{-n}^{I} \alpha_{n}^{I}
 \comma \quad 
 N_{\rm f} = \sum_{n=1}^{\infty} n S_{-n}^{a} S_{n}^{a}
 \period
\eqe

The supersymmetric boundary state $\vert  \calB    \ket$ 
is defined to preserve half the supercharges, 
\eqb\label{QBS}
    ( Q^{a} + i M_{ab}\Qtil^{b} )  \vert   \calB   \ket =
    ( Q^{\adot} + i M_{\adot\bdot}\Qtil^{\bdot} )   \vert  \calB   \ket = 0
    \period
\eqe
One finds that these conditions are satisfied by  
\eqb\label{susyBS}
   \vert  \calB   \ket  \Eqn{=} 
   \calN_{\calB} \prod_{n=1}^{\infty}\exp\biggl[
    \frac{1}{n} M_{IJ}\alpha_{-n}^{I} \alphatil_{-n}^{J}
  -iM_{ab} S_{-n}^{a} \Stil_{-n}^{b}  
 \biggr]
  \vert  \calB   \ket_{0} 
  \period
\eqe
Here, $\calN_{\calB}$ is  the  normalization constant.
The zero-mode part
$\vert  \calB   \ket_{0} = \vert  \calB    \ket_{\rm b0} \vert  \calB    \ket_{\rm f0}$ 
 is annihilated by all the positive modes and 
given by 
\eqb\label{B0} 
    \vert  \calB    \ket_{\rm b0}  
    \Eqn{=} \sum \vert k^{I} ,  k^{K} M_{KJ}  \ket
    \comma \nn \\
    \vert  \calB   \ket_{\rm f0} \Eqn{=}
   M_{IJ} \vert I\ket_{R} \vert J \ket_{L} - iM_{\adot\bdot}
    \vert \adot \ket_{R}\vert \bdot \ket_{L} \period
\eqe
The summation symbol stands
for the summation/integral over possible zero modes 
with appropriate weight. 
The bosonic zero modes  act on the oscillator vacuum as 
$\alpha_{0}^{K} \vert k^{I}, \ktil^{J} \ket = (k^{K}/2)  \vert k^{I}, \ktil^{J} \ket$,
 $\alphatil_{0}^{K} \vert k^{I}, \ktil^{J} \ket = (\ktil^{K}/2)  \vert k^{I}, \ktil^{J} \ket$, 
whereas the spinor zero modes as 
\eqb\label{S0action}
   S_{0}^{a} \vert \adot \ket_{R} 
   = \frac{1}{\sqrt{2}} \sigmabar^{I}_{\adot a} \vert I \ket_{R}
   \comma \quad 
  S_{0}^{a} \vert I \ket_{R} = \frac{1}{\sqrt{2}} \sigma^{I}_{a \adot} \vert \adot \ket_{R}
  \comma
\eqe
for the right movers and similarly for the left movers. The  matrices 
$( M_{IJ}, M_{ab}, M_{\adot\bdot} )$ 
are taken to be orthogonal ones,  
\eqb\label{Ms} 
    M_{KL} = \exp\Bigl[ \Omega_{IJ} \Sigma^{IJ}\Bigr]_{KL}  
    \comma \quad
    M_{\alpha\beta} = \exp\Bigl[ \half \Omega_{IJ} \gamma^{IJ} \Bigr]_{\alpha\beta}
    = \biggl(\begin{array}{cc} M_{ab} & 0 \\0 & M_{\adot\bdot}\end{array}\biggr)
    \comma
\eqe 
where 
$(\Sigma^{IJ})_{KL} = \delta^{I}_{K} \delta^{J}_{L} -  \delta^{J}_{K} \delta^{I}_{L}$,
$\gamma^{IJ} = (\gamma^{I} \gamma^{J} - \gamma^{J} \gamma^{I} )/2 $
and $\Omega_{IJ} = - \Omega_{JI}$ are parameters.
These SO(8) matrices  are related  by
\eqb\label{gammaM}
   \gamma^{K}M_{KI} = M \gamma^{I} M^{T}
  \comma
\eqe    
which reads in terms of the $8 \times 8$ matrices  
$
   \sigma_{a\adot}^{K} M_{KI} = M_{ab} \sigma^{I}_{b\bdot} M_{\adot\bdot} \period
$
On the boundary state, the modes of the fields satisfy the boundary conditions, 
\eqb\label{modeBS}
  ( \alpha_{n}^{I} - M_{IJ} \alphatil_{-n}^{J} )   \vert  \calB   \ket 
  = ( S^{a}_{n} + iM_{ab} \Stil_{-n}^{b} )   \vert  \calB   \ket = 0 \period
\eqe 
These are translated into  the conditions on the fields at $\tau=0$
by using the mode expansions
$\del_{-} X^{AI} = \sum \alpha_{n}^{AI} e^{-2in (\tau-\sigma)}$, 
$ S^{Aa} = \sum S_{n}^{Aa} e^{-2in (\tau-\sigma)}$ and similar  ones
for the left movers.
Successively acting on the boundary state
with the combinations of the supercharges and modes in 
(\ref{QBS}) and (\ref{modeBS}) yields consistency
conditions. One can check that these are satisfied by the orthogonality 
of the matrices, the relation (\ref{gammaM}) and the constraint 
$P^{-} = \Ptil^{-}$ which follows
from  (\ref{lcVirasoro}).

A simple example of the supersymmetric boundary state is given by setting
$M_{IJ}$ and $M_{\alpha\beta}$ to be 
\eqb\label{Mp}
   M^{(p)}_{IJ} = \biggl(\begin{array}{cc}  -\One_{p+1} & 0 \\0 & \One_{7-p} \end{array}\biggr)
   \comma \qquad M^{(p)}_{\alpha\beta} = \gamma^{1} \cdots \gamma^{p+1} \comma
\eqe
where $\One_{n}$ is the $n \times n$ unit matrix.
This corresponds to the Neumann conditions (in the open string channel) 
for $I = 1, ..., p+1$ and
the Dirichlet conditions for  $I = p+2, ..., 8$. Furthermore, in light-cone gauge,
it follows that $\del_{\sigma} X^{\pm} =0  $ from the gauge fixing condition 
and the constraints (\ref{lcVirasoro}) . This means that one also has the Dirichlet conditions 
in the light-cone directions. The boundary state thus represents the $(p+1)$-instanton,
which is related to the usual D$p$-brane by a double Wick rotation.
Keeping this relation in mind, we use the terminology ``D-brane'' 
also for the boundary state in this paper.
One can check that the coupling of the massless closed string modes
to the boundary state 
agrees with that to the (Wick rotated) black $p$-brane. 
General supersymmetric boundary states are obtained by SO(8) transformations
of $M^{(p)}_{IJ}$ and $M^{(p)}_{\alpha\beta}$. In particular, 
the sign of $M_{\alpha\beta}$ is flipped by a $2\pi$-rotation 
in all directions, which transforms a BPS state to an anti-BPS state.
Note also that the forms of the matrices in  (\ref{Mp}) are compatible 
with (\ref{Ms}) only when $p$ is odd.

\subsection{Supersymmetric boundary states in type IIA theory}

One can similarly construct the supersymmetric boundary state in type IIA theory
by flipping the chirality for the left movers, e.g., $\Stil^{a} \to \Stil^{\adot}$ and 
$\vert \adot \ket_{L} \to \vert a \ket_{L}$. 
In this case, the boundary conditions 
for  the supercharges become
\eqb\label{QBIIA}
    ( Q^{a} + i M_{a\bdot}\Qtil^{\bdot} )   \vert  \calB   \ket =
    ( Q^{\adot} + i M_{\adot b}\Qtil^{b} )   \vert  \calB   \ket = 0 \comma
\eqe
where $\Qtil^{\adot} = \sqrt{2p^{+}} \Stil^{\adot}$ are the supercharges for the 
linearly realized supersymmetry and $\Qtil^{a}$, which are defined similarly to 
(\ref{NLQ}), are those for the non-linearly realized supersymmetry.
The SO(8) matrices in (\ref{Ms}) 
are multiplied by matrices generating reflections. 
The resultant  matrices satisfy the relation (\ref{gammaM}) as before.
For instance, 
for the usual D$p$-brane with even $p$, 
one has $M^{(p)}_{IJ}$ in (\ref{Mp}) with even $p$ and
\eqb\label{MpIIA}
    M^{(p)}_{\alpha\beta} = \gamma^{9}\gamma^{1} \cdots \gamma^{p+1} 
    = \biggl(\begin{array}{cc} 0 & M_{a\bdot} \\ M_{\adot b}& 0 \end{array}\biggr)
    \period
\eqe
The boundary state in type IIA theory then takes the form which 
is obtained from  (\ref{susyBS}) and  (\ref{B0}) by replacing 
$\Stil_{n}^{a}, (M_{ab}, M_{\adot\bdot})$
and $\vert \bdot \ket_{L}$
with $\Stil_{n}^{\adot}, (M_{a\bdot}, M_{\adot b}) $ and 
$\vert b \ket_{L}$, respectively.

\subsection{Unfolding procedure}

\begin{figure}[t]
\begin{center}
\resizebox{110mm}{!}{\includegraphics{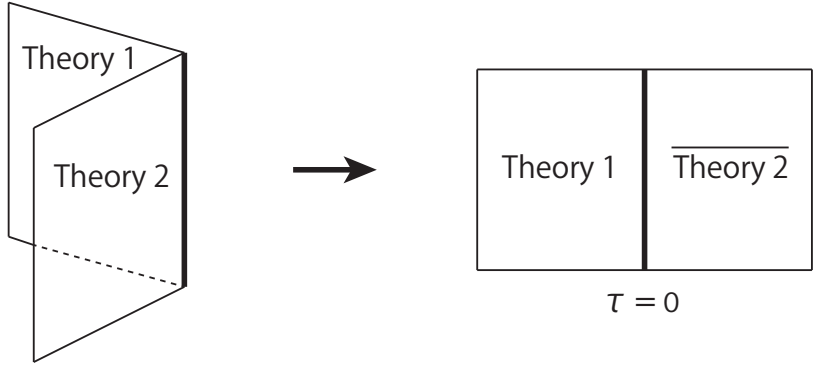}}
\end{center}
\caption{Unfolding procedure. A boundary state in a tensor product theory is
unfolded to an interface gluing theory 1 and 2.}
\label{fig:unfold}
\end{figure}

In two-dimensional CFT,  
a way to construct conformal interfaces is to unfold conformal
boundary states. Suppose that one has a boundary state 
$ \vert  \calB   \ket_{\rm CFT} 
= \sum_{i,j} c_{ij}\vert   \calB_{i} \ket_{1}  \otimes \vert  \calB_{j} \ket_{2} $
in a tensor product theory CFT$_{1}\otimes$CFT$_{2}$, which
satisfies
\eqb
   0=( L^{1}_{n} +  L^{2}_{n}- \Ltil^{1}_{-n}  -\Ltil^{2}_{-n})   \vert  \calB   \ket_{\rm CFT}
   \period
\eqe
Here,    $c_{ij}$ are  coefficients and 
$L_{n}^{A}, \Ltil_{n}^{A}$ $(A=1,2)$ are the Virasoro generators for the 
right and left movers in CFT$_{ A}$, respectively. 
Then, one can obtain  
a conformal interface gluing  CFT$_{1}$ and CFT$_{2}$ 
by unfolding the boundary state as 
\eqb
  \bbI   =  \sum_{i,j} c_{ij} \vert   \calB_{i} \ket_{1}  \cdot 
  {}_{2}\overline{ \bra  \calB_{j}  \vert } \comma
\eqe
where ${}_{2}\overline{ \bra   \calB_{j}   \vert }$ is obtained 
from  $\vert  \calB_{j}   \ket_{2}$
by the hermitian conjugation 
followed by the sign flip of the world-sheet coordinate  $\tau \to - \tau$. 
(See Figure \ref{fig:unfold}.)
The resultant interface indeed preserves the conformal invariance,
\eqb
    ( L^{1}_{n} - \Ltil^{1}_{-n} )   \bbI   =    \bbI    
     ( L^{2}_{n}- \Ltil^{2}_{-n} )  \comma
\eqe
which also means the conservation of energy across the world-sheet 
interface/defect $\bbI$.
In this construction, the interface is located at $\tau = 0$ in the world-sheet.

\section{Supersymmetric interfaces}

As we observed in the previous section, the supersymmetric boundary state
represents the D-brane and is regarded as
an analog of the conformal boundary state. 
 In this section, we construct the supersymmetric interface 
by unfolding the supersymmetric boundary state,
similarly to   the conformal  interface.

\subsection{Case of type IIB theory}

To be specific, in this subsection we consider the world-sheet interface which 
glues  two type IIB theories residing on the left and the right side of the interface,
respectively (IIB-IIB case). 
The interface is defined to satisfy the conditions on the supercharges,
\eqb
  ( Q_{1}^{a} + i R^{1}_{ab} \Qtil_{1}^{b} )   \bbI  
   \Eqn{=}  \bbI    ( Q_{2}^{a} + i R^{2}_{ab} \Qtil_{2}^{b} ) 
   \comma \label{QI1} \\
   (Q_{1}^{\adot} + i R^{1}_{\adot\bdot} \Qtil_{1}^{\bdot})  \bbI  
   \Eqn{=}  \bbI    (Q_{2}^{\adot} + i R^{2}_{\adot\bdot} \Qtil_{2}^{\bdot} ) \comma
    \label{QI2} 
\eqe
for some $R^{A}_{ab}, R^{A}_{\adot\bdot}$ $(A=1,2)$.
According to the unfolding procedure, 
we first 
double the fields, and denote the resultant modes by
\eqb
  (\alpha_{n}^{AI}, \alphatil_{n}^{AI}) \comma \quad
  (S^{Aa}_{n}, \Stil^{Aa}_{n})  \period
\eqe
We remark that we have doubled the fields just as an intermediate
step for the construction.
Then, one may consider a boundary state in which 
the bilinear forms of the oscillators are given by 
$\S_{AB}^{IJ} \alpha_{-n}^{AI} \alphatil_{-n}^{BJ} $ and $ \S_{AB}^{ab} 
S_{-n}^{Aa} \Stil_{-n}^{Bb}$,
where $\S_{AB}$'s are the ``S-matrix'' which determines the boundary conditions 
of the modes. Next, by unfolding the sector with $A=2$, 
one finds that the oscillators 
are transformed 
as $(\alpha_{n}^{2I}, \alphatil_{n}^{2I}) \to (-\alphatil_{-n}^{2I}, -\alpha_{-n}^{2I})$
and $(S^{2a}_{n}, \Stil^{2a}_{n} )\to (\Stil^{2a}_{-n}, S^{2a}_{-n})$.
This results in an interface of the form
\eqb\label{susyI}
    \bbI     \Eqn{=}   \calN_{\bbI}  \,  \bbI_{\rm b} \cdot \bbI_{\rm f} \comma \nn \\
    && \bbI_{\rm b} = 
    \prod_{n=1}^{\infty} \exp\biggl[ \frac{1}{n} \beta_{-n}^{AI} \cdot \S_{AB}^{IJ} \cdot 
   \betatil_{-n}^{BJ} \biggr]  \cdot \bbI_{\rm b0} \comma \\
   && \bbI_{\rm f} = 
    \prod_{n=1}^{\infty} \exp\biggl[-i T_{-n}^{Aa} \ast \S_{AB}^{ab} \ast 
   \Ttil_{-n}^{Bb} \biggr]  \cdot \bbI_{\rm f0} \period \nn 
\eqe
Here, the oscillators are defined by
\eqb
    && \beta_{n}^{AI} := (\alpha_{n}^{1I}, -\alphatil_{-n}^{2I})
     \comma \quad  
      \betatil_{n}^{AI} := (\alphatil_{n}^{1I}, -\alpha_{-n}^{2I})  \comma \nn \\
      && T_{n}^{Aa} := (S_{n}^{1a}, - \Stil_{-n}^{2a})
     \comma \quad 
      \Ttil_{n}^{Aa} := (\Stil_{n}^{1a}, - S_{-n}^{2a}) \comma
\eqe
and the product  $\ast$ by
\eqb
   U_{A} \ast V_{A} := \eta_{AB} U_{A}  V_{B} \comma
\eqe
with $\eta_{AB} =\diag(+1,-1)$.  
(We do not  raise or lower the indices $A,B$ by $\eta_{AB}$.)
$\calN_{\bbI}$ is the normalization constant. 
It is also understood that  the annihilation operators, or the oscillators with $A=2$, 
act  on the interface 
implicitly from the right, e.g., 
\eqb
 \exp \Bigl[ \alpha_{-n}^{1I} \alpha_{n}^{2J} \Bigr]  \cdot    \bbI     
 = \sum_{l} \frac{1}{l!} (\alpha_{-n}^{1I})^{l} \cdot  \bbI 
 \cdot  (\alpha_{n}^{2J})^{l}
 \period
\eqe
Starting from our ansatz of the form of the interface (\ref{susyI}), we would like to 
determine $S_{AB}^{IJ}$,
$S_{AB}^{ab}$  and the zero-mode factors $\bbI_{\rm b0},    \bbI_{\rm f0}$,
as well as $R^{A}_{ab}$, $R^{A}_{\adot\bdot}$ in (\ref{QI1}), (\ref{QI2}).

For this purpose, 
we first note that the oscillators  satisfy the continuity 
conditions on the interface, $  \beta_{n}^{AI}  \approx  \S_{AB}^{IJ} \betatil_{-n}^{BJ} $, 
$ \betatil_{n}^{AI}  \approx \beta_{-n}^{BJ} \S_{BA}^{JI} $ and 
$ T_{n}^{Aa}  \approx  -i \S_{AB}^{ab} \ast  \Ttil_{-n}^{Bb} $, 
$ \Ttil_{n}^{Aa}  \approx +i T_{-n}^{Bb} \ast \S_{BA}^{ba} $
for $ n \geq 1 $. 
The symbol $\approx$ denotes the relations which hold on the interface.
For example,
$ \beta_{n}^{1I}  \approx  \S_{1B}^{IJ} \betatil_{-n}^{BJ} $ stands for
$ (\alpha_{n}^{1I} - \S_{11}^{IJ} \alphatil_{-n}^{1J})    \bbI  =   \bbI   
(-\S_{12}^{IJ} \alpha^{2J}_{n})$. Next, we require that all the modes 
with  $n \in \bbZ $
have the same transformations
so that the continuity conditions give linear transformations
of the fields. This leads to the condition that $S_{AB}^{IJ}$ is orthogonal
and $S_{AB}^{ab}$ is pseudo-orthogonal, 
\eqb
   \S_{AB}^{IJ} \S_{AC}^{IK} =  \delta^{JK} \delta_{BC}
   \comma \quad 
   \S_{AB}^{ab} \ast \S_{AC}^{ac} = \delta^{bc}\eta_{BC} 
      \period
\eqe
The continuity conditions on the oscillators are then summarized as
\eqb\label{ContCond}
   \beta_{n}^{AI}  \approx  \S_{AB}^{IJ} \betatil_{-n}^{BJ} 
   \comma \quad 
   T_{n}^{Aa}  \approx  -i \S_{AB}^{ab} \ast  \Ttil_{-n}^{Bb} \comma
\eqe
for $  n \neq 0$.
Furthermore, 
in order for $(\alpha_{0}^{AI}, \alphatil_{0}^{AI})$  
to have the same transformations as the non-zero modes,
the bosonic 
zero-mode factor $\bbI_{\rm b0}$ should be of  the form,   
\eqb
  \bbI_{\rm b0} = 
     \sum 
      \vert k^{1I}, k^{BI'} \S^{I'J}_{B1} \ket _{1} \cdot 
    {}_{2}\bra -k^{2K},  -   k^{BI'} \S^{I'L}_{B2}\vert
        \period
\eqe
On the dual vacuum $(\alpha_{0}^{I}, \alphatil_{0}^{I})$ act as
$ \bra \ktil^{K}, k^{L} \vert \alpha_{0}^{I} = \bra \ktil^{K}, k^{L} \vert  (k^{I}/2)$
and 
 $ \bra \ktil^{K}, k^{L} \vert \alphatil_{0}^{I} = \bra \ktil^{K}, k^{L} \vert  (\ktil^{I}/2)$.

Now, let us impose the conditions on the supercharges (\ref{QI1}), (\ref{QI2}).
Since the linearly realized supercharges $Q_{A}^{a}$ are nothing but 
the spinor zero-modes, 
only the spinor zero-mode factor $\bbI_{\rm f0} $ is relevant for 
the conditions on $Q_{A}^{a}$. Its  general form is given by
\eqb\label{If0gen}
    \bbI_{\rm f0}   = M_{ijkl}   \vert  i \ket_{1R} \vert j \ket_{1L} 
   \cdot {}_{2L} \bra k \vert {}_{2R} \bra l \vert
   \comma
\eqe
where 
$i= ( I,\adot),    j= (J,\bdot),   k = (K,\dotc),   l=(L,\ddot)$. Assuming 
that $\bbI_{\rm f0}$ is bosonic, the coefficients $M_{ijkl}$ are  non-vanishing
only when an even number of the indices takes the vector/spinor indices.
Given the form (\ref{If0gen}) and the action of the spinor zero modes
(\ref{S0action}), the conditions (\ref{QI1}) are translated
into those for $M_{ijkl}$ and $R^{A}_{ab}$. We list them in the appendix.

One can find two simple classes of the solutions. 
In both classes, one has 
\eqb\label{Rab}
     R^{A}_{ab} = \eta_{A} M^{A}_{ab} \comma \quad \eta_{A} = \pm 1 \period
\eqe
The non-vanishing coefficients $M_{ijkl}$ in one class are given by
\eqb\label{MijklFD}
    && M^{\rm FD}_{ijkl}  = N^{1}_{ij} N^{2}_{kl} \comma \\
   && \qquad N^{1}_{IJ} = M^{1}_{IJ} \comma \ \
    N^{1}_{\adot\bdot} = -i  \eta_{1} M^{1}_{\adot\bdot} \comma \ \
    N^{2}_{KL} =  M^{2}_{LK} \comma \ \
    N^{2}_{\dotc\ddot}  = -i  \eta_{2} M^{2}_{\ddot \dotc} \comma \nn
\eqe
and in the other by
\eqb\label{MijklTP}
    && M^{\rm TP}_{ijkl} = N^{\rm id}_{il} N^{\rm rot}_{jk} \comma \\
    && \qquad N^{\rm id}_{IL} = \delta_{IL} \comma \  \
    N^{\rm id}_{\adot\ddot} =   \delta_{\adot\ddot} \comma \ \
    N^{\rm rot}_{JK} =  M^{1}_{IJ}M^{2}_{IK} \comma \ \
    N^{\rm rot}_{\bdot \dotc}  
    =  \eta_{1}\eta_{2} M^{1}_{\adot\bdot}M^{2}_{\adot \dotc} 
    \comma \nn
\eqe
where $(M_{IJ}^{A}, M_{ab}^{A}, M_{\adot\bdot}^{A})$
are sets of SO(8) matrices satisfying (\ref{gammaM}).
We have also absorbed overall constants into the normalization 
constant $\calN_{\bbI}$.
Since $\eta_{A}$ can be absorbed by $2\pi$-rotations in each 
sector with $A=1$ or $A=2$, we set $\eta_{A}= +1$ in the following.

Let us next discuss  the conditions   for the non-linearly realized 
supercharges $Q_{A}^{\adot}$.  
A way to obtain a sufficient condition for (\ref{QI2})  to hold
is as follows. First,  decompose the summation $\sum_{n \in \bbZ}$
as $ \sum_{n \geq 1} + \sum_{n > 1}$ by flipping the sign of $n$  for $n<0$.
Next, applying  (\ref{ContCond})
for $n \geq 0$, one obtains an expression in terms of 
$(\beta_{-n}^{AI}, \betatil_{-n}^{AI})$
and $(T_{-n}^{Aa}, \Ttil_{-n}^{Aa})$ with $n \geq 0$.
Requiring each term, e.g., of the form $S_{-n}^{1a} \alphatil^{1J}_{-n}$,  
to vanish gives 
a set of equations for $ \S_{AB}^{IJ}, \S_{AB}^{ab}$ and $R^{A}_{\adot\bdot}$.
We list them in the appendix. One then finds that those equations are
solved by
\eqb\label{SAB}
   && R_{\adot\bdot}^{A} = \ep_{A} M^{A}_{\adot\bdot} \comma \quad 
   \ep_{A} = \pm 1 \comma
    \\
   && \S_{AB}^{IJ}  
    =  \Biggl(\begin{array}{cc}
    a_{11} M^{1}_{IJ} 
    & a_{12} \delta_{IJ} \\
    a_{21} M^{1}_{KJ} M^{2}_{KI}
     & a_{22} M^{2}_{JI}
     \end{array}\Biggr)
     \comma \ \
      \S_{AB}^{ab}  
      = \Biggl(\begin{array}{cc} \ep_{1} a_{11} M^{1}_{ab} &
          -i a_{12} \delta_{ab}  \\
            i \ep_{1} \ep_{2}a_{21} M^{1}_{cb} M^{2}_{ca} 
            & \ep_{2}a_{22} M^{2}_{ba}
            \end{array}\Biggr)
      \comma \nn
\eqe
where $a_{AB}$ is an orthogonal matrix to maintain the (pseudo-)orthogonality
of $\S_{AB}^{IJ}$ ($\S_{AB}^{ab}$), and 
$(M_{IJ}^{A}, M_{ab}^{A}, M_{\adot\bdot}^{A})$
are sets of SO(8) matrices satisfying (\ref{gammaM}).

We still have to check some consistency conditions.
First,  both (\ref{ContCond}) with $n=0$ and (\ref{QI1}) 
give the continuity conditions  on $(S_{0}^{Aa}, \Stil_{0}^{Aa})$, 
which should be compatible.
Indeed, if $ ( Q_{1}^{a} + i M^{1}_{ab} \Qtil_{1}^{b} ) - 
( Q_{2}^{a} + i M^{2}_{ab} \Qtil_{2}^{b} )  $ is
evaluated by using (\ref{ContCond}) and (\ref{SAB}),
one finds that it vanishes on the interface  only when
$\ep_{1} a_{11} = 1+ \ep_{1} \ep_{2} a_{21} $ and $\ep_{2} a_{22} = 1 + a_{12}$.
This means that $a_{AB}$ should be either of 
\eqb\label{aAB}
     a_{AB}^{\rm FD} =  \biggl(\begin{array}{cc} \ep_{1} & 0 \\0 
     & \ep_{2} \end{array}\biggr)
     \comma \qquad 
    a_{AB}^{\rm TP} =  \biggl(\begin{array}{cc} 0 
    & -1 \\-\ep_{1} \ep_{2} & 0\end{array}\biggr)
    \period
\eqe
One can  confirm that  the former gives the same conditions on 
$(S_{0}^{Aa}, \Stil_{0}^{Aa})$
as those from $M^{\rm FD}_{ijkl}$ and the latter as from $M^{\rm TP}_{ijkl}$, 
under the identification of the SO(8) matrices 
in (\ref{MijklFD}), (\ref{MijklTP}) and those in (\ref{SAB}).
Second,  one has further conditions by successively acting on the interface  
with the combinations of the supercharges and the modes in 
(\ref{QI1}), (\ref{QI2}) and (\ref{ContCond}). 
These are checked by using (\ref{SAB}) 
and the constraint (\ref{lcVirasoro}).
Since the signs $\ep_{A}$ can be absorbed by the redefinitions 
$ (\ep_{A} M^{A}_{IJ}, M^{A}_{ab}, \ep_{A}M^{A}_{\adot\bdot})$
 $\to$ $(M^{A}_{IJ}, M^{A}_{ab}, M^{A}_{\adot\bdot})$  and 
$\ep_{1} \ep_{2}\calN_{\bbI} \to \calN_{\bbI}$, we set $\ep_{A } =+1$ in the following.

In summary, to construct the interfaces satisfying the supersymmetric conditions 
(\ref{QI1}) and  (\ref{QI2}),  we started with the 
ansatz (\ref{susyI}), which follows from 
the supersymmetric boundary state and the unfolding procedure.
 We further required 
the interfaces to induce linear transformations of the fields, which 
in particular leads to the condition that both zero and non-zero modes
transform homogeneously.
We then found  the two classes of the supersymmetric interfaces,
which we labeled by FD and TP, respectively. 

\par\bigskip
\noindent
{\it Factorized D-branes}
\par\smallskip

In one class,  the interface takes the form 
\eqb\label{IFD}
   \bbI^{\rm FD}   \Eqn{=} \calN_{\rm FD} 
       \prod_{n=1}^{\infty} e^{\frac{1}{n} M^{1}_{IJ} \alpha_{-n}^{1I}   
   \alphatil_{-n}^{1J} - i M^{1}_{ab} S_{-n}^{1a}  \Stil^{1b}_{-n} }
   \cdot \bbI_{\rm b0}^{\rm FD}   \bbI_{\rm f0}^{\rm FD} \cdot 
    \prod_{n=1}^{\infty} e^{\frac{1}{n} M^{2}_{IJ} \alpha_{n}^{2I} \alphatil_{n}^{2J}   
    + i M^{2}_{ab}  S^{2a}_{n} \Stil_{n}^{2b}  } \comma \nn \\
   && \bbI_{\rm b0}^{\rm FD} 
      = \sum 
       \vert k^{1I} ,  \ktil^{1J} \ket \bra \ktil^{2K} ,  k^{2L} \vert  
     \comma \quad 
     k^{AI} =   M_{IJ}^{A} \ktil^{AJ} \comma  
     \\ 
     &&  \bbI^{\rm FD}_{\rm f0} = \bigl(   M^{1}_{IJ} \vert I \ket \vert J \ket 
    -i M^{1}_{\adot\bdot} \vert \adot \ket \vert \bdot \ket    \bigr)
    \bigl(   \bra K \vert \bra L \vert  M^{2}_{LK} 
    -i  M^{2}_{\ddot \dotc} \bra \dotc \vert \bra  \ddot \vert      \bigr) \comma \nn
\eqe
where we have omitted the subscripts for the oscillator vacua, and 
the index $A$ has not been summed.
The supercharges and the fields satisfy the continuity conditions, 
\eqb
    \begin{array}{ll}
    0  \approx Q_{A}^{a} + i M^{A}_{ab} \Qtil_{A}^{b} 
    \comma & \quad 
    0  \approx  Q_{A}^{\adot} + i M^{A}_{\adot\bdot} \Qtil_{A}^{\adot} 
      \comma  \\
    0  \approx  S^{Aa} +  i  M^{A}_{ab} \Stil^{Ab} \comma 
    & \quad 
     0  \approx  \del_{-} X^{AI} -   M^{A}_{IJ} \del_{+} \Xtil^{AJ}
     \period
    \end{array}
\eqe
Here, we have used  the fact that the interface is at $\tau=0$.
One also finds  that the energy does not flow across the interface $\bbI^{\rm FD}$, namely,
\eqb
      L_{n}^{A} - \Ltil_{-n}^{A} \approx 0 \comma
\eqe
for $A=1,2$, where $L_{n}^{A} = L_{n}^{{\rm b} A} + L_{n}^{{\rm f} A} $, 
\eqb
    L_{n}^{{\rm b} A} = \half \sum_{m \in \bbZ} \alpha^{AI}_{n-m} \alpha^{AI}_{m}
    \comma \quad 
     L_{n}^{{\rm f}A} = \half  \sum_{m \in \bbZ} (m-\frac{n}{2}) 
    S^{Aa}_{n-m} S^{Aa}_{m} \comma 
\eqe
and similarly for the left movers.
The interface $\bbI^{\rm FD}$ is thus understood as a factorized 
D-branes, a factor of which with $A=2$ is in a conjugate form, 
``$\vert Dp_{1} \ket\bra Dp_{2} \vert $''.  
This class is regarded as an analog of the totally reflecting case of 
the  permeable conformal interfaces \cite{Bachas:2001vj}.
In the following, we call this class/case the FD class/case.

\par\bigskip
\noindent
{\it Analog of topological interfaces}
\par\smallskip

In the other class, the  interface takes the form
\eqb\label{ITP}
   \bbI^{\rm TP}   \Eqn{=} \calN_{\rm TP} 
       \prod_{n=1}^{\infty} e^{\frac{1}{n}  ( \alpha_{-n}^{1I} \alpha_{n}^{2I} 
        + M^{1}_{KI} M^{2}_{KJ} \alphatil_{-n}^{1I} \alphatil_{n}^{2J}  )}
         e^{S_{-n}^{1a}  S^{2a}_{n} + M^{1}_{ca} M^{2}_{cb} \Stil_{-n}^{1a} \Stil_{n}^{2b}  }
   \cdot \bbI_{\rm b0}^{\rm TP}   \bbI_{\rm f0}^{\rm TP}
    \comma  \nn \\
    && \bbI_{\rm b0}^{\rm TP} 
      = \sum
         \vert k^{1I} ,  \ktil^{1J} \ket \bra \ktil^{2K} ,  k^{2L} \vert 
       \comma \nn \\
       && 
       \qquad \quad \
     k^{1I} = k^{2I} \comma  \quad   M^{1}_{IJ} \ktil^{1J} = M^{2}_{IJ} \ktil^{2J} \comma
     \\ 
     &&    \bbI^{\rm TP}_{\rm f0} =
    \vert I \ket   T_{\bbI}   \bra I \vert 
    + \vert \adot \ket   T_{\bbI}  \bra \adot \vert  \comma \nn \\  
    && \qquad \quad \ 
     T_{\bbI} =M^{1}_{PJ} M^{2}_{PK} \vert J \ket \bra K \vert 
  +  M^{1}_{\dot{p} \bdot} M^{2}_{\dot{p}\dotc } \vert \bdot \ket \bra \dotc \vert
  \comma\nn 
\eqe 
where the oscillators with $A=2$ are understood as acting
 on the zero-mode factors
from the right side, as mentioned.
The supercharges and the fields satisfy the continuity conditions,
\eqb\label{TPtrans}
    \begin{array}{rclrcl}
    Q_{1}^{a}  \Eqn{\approx} Q_{2}^{a} 
    \comma & \quad 
    M^{1}_{ab}\Qtil_{1}^{b} \Eqn{\approx}  M^{2}_{ab}\Qtil_{2}^{b}     \comma  \\
    Q_{1}^{\adot}  \Eqn{\approx} Q_{2}^{\adot} 
    \comma & \quad 
    M^{1}_{\adot\bdot}\Qtil_{1}^{\bdot}  
    \Eqn{\approx}  M^{2}_{\adot\bdot}\Qtil_{2}^{\bdot}     \comma  \\
        S^{1a} \Eqn{\approx} S^{2a} \comma 
    & \quad 
   M^{1}_{ab}\Stil^{1b}  \Eqn{\approx}  M^{2} _{ab} \Stil^{2b} \comma \\
     \del_{-} X^{1I} \Eqn{\approx}  \del_{-} X^{2I}  \comma 
    & \quad 
    M^{1}_{IJ} \del_{+} X^{1J} \Eqn{\approx}  M^{2}_{IJ} \del_{+} X^{2J}  \period
    \end{array}
\eqe
We thus find that the interface $\bbI^{\rm TP}$ generates T-dualities. 
In addition, the continuity conditions of the Virasoro generators reads 
\eqb 
      L_{n}^{1}   \approx  L_{n}^{2} \comma \quad 
      \Ltil_{n}^{1} \approx \Ltil_{n}^{2} \comma
\eqe
which also implies that the energy is conserved across the interface.
Since the two sets of the Virasoro generators are conserved
across the interface, 
$  \bbI^{\rm TP}$  is regarded as  an analog 
of the totally transmissive or the topological case 
of the permeable conformal interfaces. 
It is known that the topological interfaces in two-dimensional  CFT
generate T-dualities \cite{Frohlich:2004ef}. The  transformations (\ref{TPtrans})
are in accord with this fact.
In the following, we call this class/case the TP class/case.

\subsection{Other cases}

Similarly, one can construct 
the supersymmetric interfaces gluing  type IIA theories (IIA-IIA case) and those gluing 
type IIB and type IIA theory (IIB-IIA case) just by appropriately changing the chirality 
of the left moving spinors in type IIA theories. 
For instance, the interface gluing type IIB theory on the left side and type IIA
theory on the right satisfies the continuity conditions,
\eqb\label{QIIIBA}
  ( Q_{1}^{a} + i R^{1}_{ab} \Qtil_{1}^{b} )    \bbI   
   \Eqn{=}   \bbI     ( Q_{2}^{a} + i R^{2}_{a\bdot} \Qtil_{2}^{\bdot} ) 
   \comma  \nn \\
   (Q_{1}^{\adot} + i R^{1}_{\adot\bdot} \Qtil_{1}^{\bdot})    \bbI   
   \Eqn{=}   \bbI     (Q_{2}^{\adot} + i R^{2}_{\adot b} \Qtil_{2}^{b} ) \period
\eqe
We again  find  two classes
of the interfaces. One is the FD class
corresponding    to the factorized D-branes, and 
the other is the TP class corresponding to  the topological interfaces in two-dimensional 
CFT. The interfaces take the form which  is obtained 
from (\ref{susyI}) by replacing 
$\Stil_{n}^{a}, (M^{2}_{ab}, M^{2}_{\adot\bdot})$
and ${}_{2L}\bra \dotc \vert$
with $\Stil_{n}^{\adot}, (M^{2}_{a\bdot}, M^{2}_{\adot b})  $ and 
${}_{2L}\bra c \vert$, respectively.
 
\section{Target-space properties}

In the following sections, we would like to study properties
of the supersymmetric interfaces constructed in the previous section.
The results   below  hold for interfaces of any  type of 
IIB-IIB, IIA-IIA and IIB-IIA, unless otherwise stated. 

To be specific, we  consider in this section the case where 
the target space is not compactified, and choose
\eqb 
    M^{A}_{IJ} = M_{IJ}^{(p_{A})} \comma
\eqe
where $M_{IJ}^{(p)}$ is given in (\ref{Mp}) with  odd or even $p$. 
Since $k^{AI} = \ktil^{AI}$ for the non-compactified target space, 
the momenta are vanishing in the  Neumann directions, i.e., 
$k^{AI} = \ktil^{AI} = 0$ for $I = 1, ..., p_{A}+1$.

\subsection{Target-space geometry}

The target-space geometry of the supersymmetric interface can be studied 
in parallel with that for the topological  interface in two-dimensional CFT 
\cite{Fuchs:2007fw}.
First, similarly to the D-brane we introduce  the position moduli.
Taking into account the allowed zero modes, 
we set $\bbI_{\rm b0}$ to be
\eqb\label{Ib0YFD}
   \bbI^{\rm FD}_{\rm b0}(Y)  = 
     \int \frac{d^{8}k^{1}}{(2\pi)^{8}} \frac{d^{8}k^{2}}{(2\pi)^{8}} 
      \, e^{-ik^{1} \cdot Y_{1}}\vert k^{1I} \ket \bra k^{2J} \vert 
       e^{ik^{2} \cdot  Y_{2}}  
     \prod_{I=1}^{p_{1}+1}  2\pi \delta(k^{1I})
     \prod_{J=1}^{p_{2}+1}  2\pi \delta(k^{2J})
     \comma
\eqe 
in the FD case, and 
\eqb\label{Ib0YTP}
   \bbI^{\rm TP}_{\rm b0}(Y)  = 
     \int \frac{d^{8}k}{(2\pi)^{8}} \, e^{-ik \cdot  Y}\vert k^{I} \ket \bra k^{J} \vert 
     \prod_{J=1}^{p+1}  2\pi \delta(k^{J})\comma
\eqe 
in the TP case. Here, we have omitted the vector indices in the contraction, 
and set 
$ \vert k^{I} \ket = \vert k^{I}, \ktil^{I} = k^{I} \ket $ and 
$ p = \max(p_{1},p_{2}) $.
To probe the target-space geometry, we further introduce  the localized states, 
\eqb
   \vert x  \ket =   \int \frac{d^{8}k}{(2\pi)^{8}} \, e^{-ik \cdot  x} \vert k^{I} \ket \period
\eqe
The amplitudes between these states in the presence of an interface is then 
given by 
\eqb
   \bra x  \vert \,  \bbI^{\rm FD}_{\rm b0} \, \vert x' \ket \Eqn{=} 
    \prod_{I=p_{1}+2}^{8} \delta(x_{I} -Y_{1}^{I})
    \prod_{I=p_{2}+2}^{8} \delta(x'_{I} -Y_{2}^{I}) \comma \nn \\
    \bra x  \vert \, \bbI^{\rm TP}_{\rm b0} \, \vert x' \ket \Eqn{=} 
    \prod_{I=p+2}^{8} \delta(x_{I} - x'_{I} -Y^{I}) \period
\eqe
The result in the FD case means that each D-brane factor in the interface
is localized at $x = Y_1$ or  $x' = Y_{2}$ 
  in the Dirichlet directions, as usual.
From the result in the TP case, we  find 
that the interface is localized in a submanifold  $x = x' +Y$ 
in the common Dirichlet directions 
in the doubled (transverse) target  space $\bbR^{8} \times \bbR^{8}  \ni (x,x')$. 
Such submanifolds have been  named ``bi-branes'' in the case of the topological
conformal interface \cite{Fuchs:2007fw}.

\subsection{Coupling through interfaces}

The bulk fields in the sector labeled by $A=1$ and those by $A=2$
couple to each other through the interface. For instance, let us 
consider the massless NS-NS fields,
\eqb
   \vert \zeta \kket := \zeta_{IJ} \vert I \ket \vert J \ket \comma
\eqe
where  we have omitted the momentum factor. The coupling is 
then read off from
\eqb\label{zetaIzeta}
  \bbra \zeta  \vert \, \bbI_{\rm f0}  \, \vert \zeta' \kket  
  = \zeta^{\ast}_{IJ} M_{IJKL} \zeta'_{LK} \period
\eqe
In the FD case, the right-hand side becomes  
$ (\zeta^{\ast}_{IJ} M_{IJ}^{(p_{1})})  (\zeta'_{LK} M_{LK}^{(p_{2})}) $, 
and each factor represents the coupling between the massless fields 
and the D$p_{1}$/D$p_{2}$-brane.
By decomposing $M_{IJ}^{(p_{A})}$ according to the SO(8) representations, 
one finds that  each factor gives  the source equations for 
the black $p_{A}$-brane at the linearized level \cite{Green:1996um}.
 
In the TP case,  the right-hand side of  (\ref{zetaIzeta}) becomes
\eqb\label{zetacoupling}
 \zeta^{\ast}_{IJ} M_{PJ}^{(p_{1})} M_{PK}^{(p_{2})} \zeta'_{IK}
= \sum_{I,J} \zeta^{\ast}_{IJ} \zeta'_{IJ} \mu_{J}
 \comma
\eqe
where $\mu_{J} = -1$
for $ p'+2 \leq J \leq p+1$ and $+1$ otherwise, and $p' = \min(p_{1},p_{2}) $.
For example, when $p=p'+1$ in the IIB-IIA case, the above coupling reads
\eqb\label{TPcouple}
    -\zeta^{\ast}_{p+1 \, p+1 } \zeta'_{p+1 \, p+1 }
    + \zeta^{\ast}_{(p+1\, I)}\zeta'_{[ p+1\, I ]} + \zeta^{\ast}_{[p+1\, I ]}\zeta'_{(p+1\, I)}
    + \sum_{I,J \neq p+1}\zeta^{\ast}_{IJ} \zeta'_{IJ} \comma
\eqe
where 
$\zeta_{(IJ)} = (\zeta_{IJ} + \zeta_{JI})/2$, 
$\zeta_{[IJ]} = (\zeta_{IJ} - \zeta_{JI})/2$.
This is in accord with the Buscher rules for the metric and the B-field,
whose  non-trivial part at the linearized level is given via
\eqb\label{BRule}
    g'_{p+1\, p+1} = \frac{1}{g_{p+1\, p+1}} \comma 
    \quad 
    g'_{p+1\, I} = \frac{b_{p+1\, I}}{g_{p+1\, p+1}}
    \comma \quad
    b'_{p+1\, I} = \frac{g_{p+1\, I}}{g_{p+1\, p+1}}
     \period
\eqe
Indeed,  $\zeta_{(IJ)}$ and $\zeta_{[IJ]}$ correspond to 
the fluctuations around the background 
$h_{IJ} = g_{IJ} -\delta_{IJ}$ and $b_{IJ}$ itself, respectively, and similarly 
for $\zeta'_{(IJ)}$, $\zeta'_{[IJ]}$. 
Thus, from (\ref{TPcouple}) one finds that $\vert \zeta \kket$ couples, or 
is continued to, $ \vert \zeta' \kket$ according to  (\ref{BRule}).
In addition, when $\zeta_{IJ} \sim \delta_{IJ}$, (\ref{zetacoupling})
shows that  the fluctuation of the dilaton $ \phi - \phi_{0}= h_{II}/4$ 
couples to  
$\phi' -\phi_{0}= h'_{II}/4 = (h_{II}-2 h_{p+1 \, p+1})/4$.
This also agrees with
the Buscher rule for the dilaton $ \phi' = \phi - \frac{1}{2}\log g_{p+1 \, p+1}$.

\section{Partition functions with interfaces inserted}

Next, let us consider the partition functions with the interfaces 
inserted, from which one can read off the spectrum 
of the modes coupled to the interfaces and the Casimir energy between them.
Here, we follow similar computations in 
\cite{Bachas:2001vj,Bachas:2007td,Sakai:2008tt}.

To be concrete, we consider the partition function where  
a pair of an interface and its conjugate is inserted.
As in the case of the D-brane \cite{Polchinski:1987tu}, 
the conjugate interface $\overline{\bbI}$ 
is defined by a CPT conjugation of $\bbI$,  which consists of  hermitian
 conjugation, complex conjugation of c-numbers and a $\pi$-rotation. 
  Consequently, one has
 $\overline{\bbI} = 
 \calN_{\bbI}   \overline{\bbI}_{\rm b} \cdot \overline{\bbI}_{\rm f} $, 
 where 
\eqb
  \begin{array}{rclrcl}
     \overline{\bbI}_{\rm b} \Eqn{=} 
    \prod_{n=1}^{\infty} e^{\frac{1}{n} \betatil_{n}^{BJ} \cdot \S_{AB}^{IJ} 
    \cdot  \beta_{n}^{AI} }  \cdot \overline{\bbI}_{\rm b0} \comma \quad 
    & && \\
   \overline{\bbI}_{\rm f} \Eqn{=} 
    \prod_{n=1}^{\infty} e^{-i  \Ttil_{n}^{Bb}\ast \S_{AB}^{ab} \ast 
    T_{n}^{Aa} } \cdot \overline{\bbI}_{\rm f0} \comma  
    &
    \overline{\bbI}_{\rm f0} 
    \Eqn{=}  M_{ijkl}   \vert  l \ket_{2R} \vert k \ket_{2L} 
   \cdot {}_{1L} \bra j \vert {}_{1R} \bra i \vert \comma
   \end{array}
\eqe
and $\overline{\bbI}_{\rm b0} =  \bbI^{\dagger}_{\rm b0}$
for the bosonic zero-mode factors of the types in (\ref{Ib0YFD}), (\ref{Ib0YTP}).
When  IIA theory is involved, the index structure of the spinor part 
should be modified appropriately.
Then, the partition function in question is given by
\eqb\label{ZIIbar}
    Z \Eqn{=} \tr \Bigl(    \bbI   q_{2} ^{L_{0}^{2}+\Ltil_{0}^{2}}   \overline{\bbI}   
       q_{1} ^{L_{0}^{1}+\Ltil_{0}^{1}} \Bigr) 
    =: \calN_{\bbI}^{2}   Z_{\rm b}^{\rm osc}   Z_{\rm b0} 
        Z_{\rm f}^{\rm osc}  Z_{\rm f0} \period
\eqe
Here,  $q_{i} = e^{-2 \pi t_{i}}$ $(i=1,2)$ are  
parameters, $Z_{\rm b/f}^{\rm osc}$  is the contribution
from the bosonic/spinor oscillators, and $Z_{\rm b0/f0} $
is that from the bosonic/spinor zero-mode factors, 
\eqb
   Z_{\rm b0} \Eqn{=} \tr \Bigl(   \bbI_{\rm b0}   q_{2}^{L_{0}^{2}+\Ltil_{0}^{2}}
        \overline{\bbI}_{\rm b0}   q_{1}^{L_{0}^{1}+\Ltil_{0}^{1}}  \Bigr) \comma \nn \\
   Z_{\rm f0} \Eqn{=} \tr \bigl(   \bbI_{\rm f0}    \overline{\bbI}_{\rm f0}  \bigr)  
   \period
\eqe

To evaluate the bosonic oscillator part, we first linearize the oscillator
bi-linear forms by  the formula
\eqb\label{eCDb}
e^{{C}  {D}} = \int_{\bbC} \frac{d^{2} {z}}{\pi} \, e^{- {z} {\zbar} 
   - {z} {C} - {\zbar}  {D}} \comma
\eqe
which hold for  bosonic operators satisfying $[C, D] = 0$. Next, we transfer 
the Virasoro generators $L_{0}^{Ab}, \Ltil_{0}^{Ab}$ 
using  $[L_{0}^{A}, \alpha^{AI}_{n}] = -n \alpha^{AI}_{n}$ until they hit the zero-mode
factor $\bbI_{\rm b0}$
or $\overline{\bbI}_{\rm b0}$. Furthermore,  by the operator identity 
$e^{C}e^{D} = e^{D}e^{C} e^{[C,D]}$ which is valid when $[C,D]$
is a c-number,  we commute the oscillators to be annihilated on the zero-mode
factors. It then follows that
\eqb\label{Zb}
   Z_{\rm b}^{\rm osc}  \Eqn{=} 
   \prod_{n=1}^{\infty} \int \frac{d^{2}{z}_{n}}{\pi^{16}} 
   \frac{d^{2}{w}_{n}}{\pi^{16}}
       e^{- {z}_{n}{\zbar}_{n} - {w}_{n} {\wbar}_{n} } 
       \cdot  
         e^{q^{n}_{1} \bigl[
          (\wbar_{n1} \S^{T}_{11}-\wbar_{n2} \S^{T}_{12}) z_{n1} 
        +  (\zbar_{n1} \S_{11} - \zbar_{n2} \S_{21}) w_{n1}  \bigr] } 
        \nn  \\
        &&  \qquad 
         \times \
       e^{q^{n}_{2} \bigl[ 
        (\wbar_{n2} \S^{T}_{22} - \wbar_{n1} \S^{T}_{21}) z_{n2} 
        +  (\zbar_{n2} \S_{22}-\zbar_{n1} \S_{12}) w_{n2}  \bigr] }   \\
        \Eqn{=}  \prod_{n=1}^{\infty} \det\!{}^{-1} D^{\rm b}_{n} \comma \nn
\eqe
where ${z}_{n} = (z_{1n}^{I}, z_{2n}^{I})$, ${w}_{n} = (w_{1n}^{I}, w_{2n}^{I})$
$(I=1, ..., 8)$ and the vector indices have been suppressed.
The matrices in the determinants  are given by
\eqb
    && D_{n}^{\rm b} =  D_{n}^{\rm 1b}  D_{n}^{\rm 2b}  D_{n}^{\rm 3b}
     \comma \nn \\
      && \quad D^{\rm 1b}_{n} = \One  -q_{1}^{n}
   (q_{1}^{n} \S^{T}_{11} \S_{11} +q_{2}^{n} \S^{T}_{21} \S_{21}) \comma  \\
  &&  \quad D^{\rm 2b}_{n} = \One -q_{2}^{n}
    (q_{2}^{n} \S^{T}_{22} \S_{22} + q_{1}^{n} \S^{T}_{12} \S_{12})
   \comma  \nn \\
 &&   \quad D^{\rm 3b}_{n} = \One
    - q_{1}^{n}q_{2}^{n} (D^{\rm 1b}_{n})^{-1}
      (q_{1}^{n} \S^{T}_{11} \S_{12}+ q_{2}^{n} \S^{T}_{21} \S_{22})
    (D^{\rm 2b}_{n})^{-1}
    (q_{2}^{n} \S^{T}_{22} \S_{21}+q_{1}^{n} \S^{T}_{12} \S_{11} ) \period \nn
\eqe

To evaluate the spinor oscillator part, we linearize the oscillator
bi-linear forms by 
\eqb
   e^{CD} = \int d\theta d\thetabar \, 
   e^{-\theta \thetabar -\theta C -\thetabar D} \comma
\eqe
which hold  for fermionic operators satisfying $\{C, D \} =0$.
Repeating similar algebras in the above, we find that
\eqb\label{Zf}
    Z_{\rm f}^{\rm osc}  \Eqn{=} 
    \prod_{n=1}^{\infty} \det D^{\rm f}_{n} \comma
\eqe
where
\eqb\label{Dfn}
    && D_{n}^{\rm f} =  D_{n}^{\rm 1f}  D_{n}^{\rm 2f}  D_{n}^{\rm 3f}
     \comma \nn \\
     && \quad  D^{\rm 1f}_{n} = \One  -q_{1}^{n}
    (q_{1}^{n} \S_{11}^{T} \S_{11} - q_{2}^{n} \S_{21}^{T} \S_{21})
   \comma   \\
   && \quad D^{\rm 2f}_{n} = \One -q_{2}^{n}
   ( q_{2}^{n} \S_{22}^{T} \S_{22}- q_{1}^{n} \S_{12}^{T} \S_{12})
   \comma \nn \\
     && \quad D^{\rm 3f}_{n} = \One  - q_{1}^{n}q_{2}^{n} (D^{\rm 2f}_{n})^{-1}
    (q_{2}^{n} \S_{22}^{T} \S_{21} -q_{1}^{n} \S_{12}^{T} \S_{11}) 
    (D^{\rm 1f}_{n})^{-1}
    ( q_{1}^{n} \S_{11}^{T} \S_{12}- q_{2}^{n} \S_{21}^{T} \S_{22})  \period \nn
\eqe

The matrices in the oscillator parts are simplified by substituting 
the generic  forms of $S_{AB}^{IJ}, S_{AB}^{ab}$ in (\ref{SAB}) without 
 using (\ref{aAB}):
\eqb\label{Dn}
    D_{n}^{\rm b} = D_{n}^{\rm f}
    \Eqn{=}  \Bigl[ 1-(q_{1}^{2n} +q_{2}^{2n}) \cos^{2}\vartheta
       - 2 q_{1}^{n} q_{2}^{n} \sin^{2} \vartheta+ q_{1}^{2n}q_{2}^{2n} \Bigr]
       \cdot \One_{8}
    \comma  
\eqe
where we have set $a_{11} = \cos \vartheta$. 
The bosonic part $ D_{n}^{\rm b}$ is a simple generalization
(eight copies)  of the result in the $c=1$ permeable 
conformal interface \cite{Bachas:2001vj}.

The evaluation of the zero-mode part is straightforward.
For example, for $\bbI_{\rm b0}^{\rm FD}$ and $\bbI_{\rm b0}^{\rm TP}$
in (\ref{Ib0YFD}) and (\ref{Ib0YTP}), we find that 
\eqb\label{Zb0}
    Z^{\rm FD}_{\rm b0}  \Eqn{=} V^{p_{1}+p_{2}+2} (\pi \sqrt{2 t_{1}} )^{p_{1}-7}
      (\pi \sqrt{2 t_{2}} )^{p_{2}-7}
    \comma \nn \\
    Z^{\rm TP}_{\rm b0}  \Eqn{=} V^{p+9} (\pi \sqrt{2 t} )^{p-7}
    \comma
\eqe
where $V = 2 \pi \delta(0)$ is the volume and  
we have set $q_{1}q_{2} = q = e^{-2\pi t}$. 
In addition, because of  the supertrace 
$\tr (\vert \adot \ket \bra \bdot \vert) = - \delta_{\adot\bdot}$,
the contributions from the NS-NS and R-R sectors cancel each other, and hence
\eqb\label{Zf0}
   Z_{\rm f0} = (\delta_{II}-\delta_{\adot \adot})^{2} = 0 \comma
\eqe
for both FD and TP cases.

In sum, the partition function with a pair of an interface and its conjugate
inserted is given by 
\eqb
   Z_{\rm FD} \Eqn{=} \calN_{\rm FD}^{2} Z_{\rm b0} Z_{\rm f0} 
   Z_{\rm b}^{\rm osc}  Z_{\rm f}^{\rm osc}  \comma \quad
   \bigl( Z_{\rm b}^{\rm osc} \bigr)^{-1} = Z_{\rm f}^{\rm osc}  
  = \prod_{n=1}^{\infty}
  (1-q_{1}^{2n})^{8}(1-q_{2}^{2n})^{8} 
   \comma
   \nn \\
    Z_{\rm TP} \Eqn{=} \calN_{\rm TP}^{2} Z_{\rm b0} Z_{\rm f0}
    Z_{\rm b}^{\rm osc}  Z_{\rm f}^{\rm osc}  \comma \quad
   \bigl( Z_{\rm b}^{\rm osc} \bigr)^{-1} = Z_{\rm f}^{\rm osc}  
    = \prod_{n=1}^{\infty} (1-q_{1}^{n}q_{2}^{n})^{16}
         \comma
\eqe
in the FD and TP cases, respectively. One can confirm 
that $Z_{\rm FD}$ corresponds to  the product of a cylinder amplitude
between D$p_{1}$-branes with the modular parameter $t_{1}$  
and the one between D$p_{2}$-branes with the modular parameter $t_{2}$
(see, e.g., \cite{Green:1996um}).
If one requires $Z_{\rm FD}$  to precisely match
the product of the D-brane amplitudes, 
the normalization constant $\calN_{\rm FD}$ is fixed.
The result in the TP case is regarded  as a square of  ordinary 
cylinder amplitudes between D-branes with the modular parameter
$t= t_{1} + t_{2}$. This is in accord with the interpretation
that $\bbI^{\rm TP}$ is ``topological'' and the modes in each sector can 
propagate (almost) freely across it.
Though the normalization constant $\calN_{\rm TP}$ would be fixed
by requiring $Z_{\rm TP}$ to match the D-brane amplitudes,
it is still an open question what condition should be imposed 
on the normalization of the interfaces gluing two different theories.
We refer to \cite{Petkova:2000ip,Fuchs:2007tx,Bachas:2007td} 
for the determination of the normalization of the topological
interface in rational or $c=1$ CFT.

In  the limit where $q_{1} \to 1$ or $q_{2} \to 1$, 
the interface and its conjugate are fused, and one may extract from 
$Z$ the spectrum of the modes which couple to  $\bbI$ or $\overline{\bbI}$.  
As is clear from the results in the above,  the spectrum 
for the supersymmetric interfaces we have constructed is essentially 
the same as that for ordinary D-branes.

On the other hand, first taking an opposite limit, e.g.,  $q_{2} \to 0$,
and then $q_{1} \to 1$, one obtains 
the Casimir energy between the interface and its conjugate. 
As in the case of the permeable conformal interface, 
one finds 
\eqb
   {\cal E} =  {\cal E}_{\rm b} + {\cal E}_{\rm f}
   \comma \quad 
   {\cal E}_{\rm b} = - {\cal E}_{\rm f} = -  \frac{1}{\pi d} \Li_{2} (\cos^{2} \vartheta) \comma
\eqe
where 
${\cal E}_{\rm b} $, ${\cal E}_{\rm f}$ are contributions from the bosons and the spinors, 
respectively, and 
$\Li_{2}(x) = \sum_{n=1}^{\infty } x^{n}/n^{2}$ is the dilogarithm function.
We have also set $q_{1} = e^{-2\pi d/T}$ with $2d$ and $2T$ being the distances
between and along the interfaces, respectively. 
This is a simple generalization of  the result in \cite{Bachas:2001vj}.
The total Casimir energy just vanishes due to  the supersymmetry.

The computation of the partition in this section can be generalized
to more general settings. First, 
when the conjugate  interface $\overline{\bbI}$ is replaced by the conjugate 
of the ``anti-interfaces''  (analog of the anti-D-branes) 
where the  signs of 
the SO(8) spinor matrices $M_{\alpha\beta}^{A}$
are different from those in $\bbI$  \cite{Bachas:2001vj},  
extra signs appear in  $Z_{\rm f0}$ and in $D_{n}^{\rm f}$ 
through $\S_{AB}^{T}$. In this case, 
the oscillator contributions $D_{n}^{\rm b}$ and 
$D_{n}^{\rm f}$  do not cancel each other anymore.
The zero-mode part also changes
depending on the signs of $M_{\alpha\beta}^{A}$. 
Second, one can insert more
interfaces into the partition function along the line of \cite{Sakai:2008tt}, 
from which, e.g., 
entanglement  entropy across the interface can be derived
by using the replica trick.

\section{Transformation of D-branes}

In two-dimensional CFT, the conformal interface transforms a set of 
conformal boundary states (D-branes) to another. Similarly, 
the supersymmetric interface transforms D-branes in type II theories.
 In this section, we derive those transformations.
To be specific, we consider the interface gluing type IIB and IIA theory.
The results for other types of IIB-IIB and IIA-IIA
follow simply by appropriately changing the chirality of the relevant spinors and
the index structure of the matrices.

In order to define the transformation of a supersymmetric boundary  state by 
an interface, we follow a similar procedure in the fusion of the $c=1$ conformal 
interfaces \cite{Bachas:2007td}.
We then regularize the product of the interface and the boundary state by
a parameter $q$, and  take the limit $q \to 1$:
\eqb\label{Dtrans}
     \vert  \calB'   \ket \Eqn{:=}   \lim_{q\to 1}   \vert  \calB'  \ket^{q} =:
       \lim_{q\to 1}  C(q)   \bbI   q^{L_{0}^{2} +\Ltil_{0}^{2}}   \vert  \calB   \ket
       \period
\eqe
Here, $\vert  \calB   \ket$ is a boundary state in type IIA theory. We denote
the SO(8) matrices appearing there by 
$(M^{\calB}_{IJ}, M^{\calB}_{a\bdot}, M^{\calB}_{\adot b}) $.
The interface $\bbI$ is gluing type IIB theory on the left and
IIA theory on the right. The constant $C(q)$, which depends on 
$\bbI$ and $\vert  \calB   \ket$, should be adjusted appropriately. 
Below, we  decompose the  regularized product
as $ \vert  \calB'   \ket^{q} 
= C(q)   \calN_{\bbI}   \calN_{\calB}   \vert \calB'\ket^{q}_{\rm b} 
\vert \calB'   \ket^{q}_{\rm f} $ 
into the normalization factors and the contributions from the bosons and 
the spinors, respectively.

The bosonic/spinor factor $\vert  \calB'  \ket^{q}_{\rm b/f}$ is evaluated
as in the previous section.
For the bosonic factor, we first linearize the bi-linear forms
in $\bbI$ and $ \vert  \calB   \ket$ using  (\ref{eCDb}). Next, 
 the oscillators and the Virasoro generators   are commuted until 
they hit the oscillator vacua to be annihilated or become c-numbers.
We then find that
\eqb
  \vert  \calB'   \ket^{q}_{\rm b} \Eqn{=} 
    \prod_{n} \int \frac{d^{2}{z}_{n}}{\pi^{16}}  \frac{d^{2}{w}_{n2}}{\pi^{8}}
       e^{- {z}_{n} {\zbar}_{n} - {w}_{n2} {\wbar}_{n2} } 
       e^{q^{n} \bigl[  \wbar_{n2}M^{\calB} z_{n2} 
       +(\zbar_{n2} \S_{22} -\zbar_{n1} \S_{12})w_{n2}  \bigr] }
       \nn \\ && 
       \qquad \times \ 
       e^{-\frac{1}{n} z_{n1} \alpha_{-n}^{1} -(\zbar_{n1} \S_{11}-\zbar_{n2} \S_{21})
       \tilalpha_{-n}^{1}}
        \cdot \vert  \calB'   \ket_{\rm b0}^{q} 
       \\
       \Eqn{=}
       \det\!{}^{-1} \calD_{\rm b}  \cdot \prod_{n}
     e^{\frac{1}{n}\alpha_{-n}^{1} (\S_{11} + q^{2n}  \S_{12} M^{\calB} 
     \calD_{\rm b}^{-1} \S_{21})
     \alphatil_{-n}^{1}} 
    \cdot \vert  \calB'   \ket_{\rm b0}^{q} \comma \nn
\eqe
where ${z}_{n} = (z_{1n}^{I}, z_{2n}^{I})$, $w_{n2} = (w_{n2}^{I})$,
$ \vert  \calB'  \ket_{\rm b0}^{q} :=  
\bbI_{\rm b0}   q^{L_{0}^{2}+\Ltil_{0}^{2}}   \vert  \calB   \ket_{\rm b0}$
and 
\eqb
     (\calD_{\rm b})_{IJ}=  (\One -q^{2n} S_{22} M^{\calB} )_{IJ} \period
\eqe
Similarly, for the spinor factor we find that
\eqb
  \vert \calB'   \ket^{q}_{\rm f} \Eqn{=} 
  \det \calD_{\rm f} \times
  \prod_{n} e^{-iS_{-n}^{1}( \S_{11}+q^{2n} \S_{12}M 
  \calD_{\rm f}^{-1} \S_{21})\Stil_{-n}^{1}}      
    \cdot \vert  \calB'   \ket_{\rm f0}\comma
\eqe     
where
\eqb
    && \vert  \calB'   \ket_{\rm f0} :=  \bbI_{\rm f0}  \vert  \calB   \ket_{\rm f0}  \\
    && \qquad =
   (M_{IJKL} M^{\calB}_{LK} - i M_{IJ c \ddot} 
   M^{\calB}_{\ddot c}) \vert I \ket \vert J \ket
   + (M_{\adot\bdot KL} M^{\calB}_{LK} - i M_{\adot\bdot c \ddot} 
   M^{\calB}_{\ddot c}) 
   \vert \adot \ket \vert \bdot \ket \comma \nn
\eqe
 and 
\eqb
   (\calD_{\rm f})_{\adot\bdot}
   = ( \One -q^{2n} \S_{22} M^{\calB})_{\adot\bdot} \period
\eqe

The results so far are generic. To compute the remaining bosonic zero-mode factor,
we specialize to the case where $\bbI_{\rm b0}$ is given by 
$\bbI_{\rm b0}^{\rm FD}$  in  (\ref{Ib0YFD}) or  $\bbI_{\rm b0}^{\rm TP}$ in (\ref{Ib0YTP}), 
and $\vert  \calB  \ket_{\rm b0}$ by
\eqb\label{Bb0Yp}
     \vert  \calB   \ket_{\rm b0}(Y_{\calB}, p_{\calB}) 
     \Eqn{=}  \int \frac{d^{8}k}{(2\pi)^{8}} \, e^{-ik \cdot Y_{\calB}} \vert k^{I} \ket
     \prod_{J=1}^{p_{\calB}+1} 2\pi \delta(k^{J}) \comma
\eqe
which corresponds to the choice $M_{IJ}^{\calB} = M_{IJ}^{(p_{\calB})}$
in a non-compactified target space.
We then find that 
\eqb
    \vert  \calB'   \ket_{\rm b0}^{q; {\rm FD}}
    \Eqn{\ \stackrel{q \to 1}{\longrightarrow} \ }  
    V^{r'+1} \prod_{J=r+2}^{8}  \delta(Y_{2}^{J} -Y_{\calB}^{J}) \times
    \vert  \calB  \ket_{\rm b0}(Y_{1},p_{1})
    \comma \nn \\
     \vert  \calB'   \ket_{\rm b0}^{q; {\rm TP}}
       \Eqn{\ \stackrel{q \to 1}{\longrightarrow} \ }  
    V^{s'+1}  \times \vert  \calB  \ket_{\rm b0}(Y + Y_{\calB}, s)
    \comma
\eqe
for $\bbI_{\rm b0}^{\rm FD}$ and $\bbI_{\rm b0}^{\rm TP}$, respectively,
where $r= \max(p_{2},p_{\calB} )$, 
$r'= \min(p_{2},p_{\calB})$, $s= \max(p, p_{\calB})$  and 
$s'= \min(p, p_{\calB} )$.

Combining all, we obtain the transformed 
supersymmetric boundary state $\vert  \calB'   \ket $.
In the FD case, we have 
\eqb
    && \vert  \calB'   \ket^{\rm FD} =
    \calN_{\calB'} 
   \prod_{n=1}^{\infty} e^{  \frac{1}{n} M^{1}_{IJ}\alpha_{-n}^{1I} \alphatil_{-n}^{1J}
  -iM^{1}_{ab} S_{-n}^{1a} \Stil_{-n}^{1b} }
  \vert  \calB'  \ket_{\rm 0}  \comma \nn \\
  && 
  \qquad \quad \vert  \calB   \ket_{\rm b0}   = \vert  \calB'  \ket_{b0}^{q=1} 
  \comma \ \
   \vert  \calB'  \ket_{\rm f0} = 
     M^{1}_{IJ}   \vert I \ket \vert J  \ket  -i  M^{1}_{\adot\bdot} 
  \vert \adot \ket \vert \bdot \ket \comma
\eqe
where
\eqb
  \calN_{\calB'} = \calN_{\bbI}   \calN_{\calB} 
    (M^{2}_{LK } M^{\calB}_{LK} -  M^{2}_{\ddot c} M^{\calB}_{\ddot c})
     \cdot \lim_{q \to 1} C(q)  
    \det\!{}^{-1}\calD_{\rm b}   \det \calD_{\rm f} \comma
\eqe
and 
\eqb
     (\calD_{\rm b})_{IJ} = \One_{IJ}-  q^{2n} M^{2}_{KI} M^{\calB}_{KJ} \comma 
     \quad (\calD_{\rm f})_{\adot\bdot} = \One_{\adot\bdot}
     -  q^{2n} M^{2}_{c\adot} M^{\calB}_{c\bdot}
   \period
\eqe
Since the factors in $\calN_{\calB'}$ and $\vert  \calB   \ket_{\rm b0} $ 
may be vanishing or diverging, the constant $C(q)$ should be
chosen accordingly, taking also into account the normalization
of the boundary state and the interface. 
The resultant boundary state essentially gives the
left-hand factor of the factorized D-branes $\bbI$, e.g.,  
the ordinary D$p_{1}$-brane when $M_{IJ}^{1} = M_{IJ}^{(p_{1})}$ . 

In the TP case, we have 
\eqb
    && \vert  \calB'  \ket^{\rm TP} =
    \calN_{\calB'} 
   \prod_{n=1}^{\infty} e^{  \frac{1}{n} M^{\calB'}_{IJ}\alpha_{-n}^{1I} 
   \alphatil_{-n}^{1J}  -iM^{\calB'}_{ab} S_{-n}^{1a} \Stil_{-n}^{1b} }
  \vert  \calB'  \ket_{\rm 0}  \comma \nn \\
  && 
  \qquad \quad \vert  \calB   \ket_{\rm b0}   = \vert  \calB'  \ket_{b0}^{q=1} 
    \comma \ \
   \vert  \calB'  \ket_{\rm f0} = 
     M^{\calB'}_{IJ}   \vert I \ket \vert J  \ket  -i  M^{\calB'}_{\adot\bdot} 
  \vert \adot \ket \vert \bdot \ket \comma
\eqe
where 
\eqb\label{TPMMM}
   M^{\calB'}_{IJ} = M^{\calB}_{IK} M^{2}_{PK} M^{1}_{PJ}
   \comma \quad   
   M^{\calB'}_{ab} = M^{\calB}_{a\bdot} M^{2}_{c\bdot} M^{1}_{cb} 
   \comma
\eqe
and 
\eqb
   \calN_{\calB'} = \calN_{\bbI}   \calN_{\calB}  \lim_{q\to1} C(q) \period
\eqe
We note $\calD_{\rm b} = \calD_{\rm f} = \One_{8}$ in this case.
The resultant boundary state describes a D-brane associated with 
the matrices $M^{\calB'} = M^{\calB} (M^{2})^{T} M^{1}$ with vector and 
spinor indices, respectively. 
For the choices of the zero-mode factors (\ref{Ib0YTP}) and (\ref{Bb0Yp}),
the position moduli are additive and allowed in the common 
Dirichlet directions for $\bbI$ and $\vert  \calB   \ket$. 

\section{Summary and discussion}

We have constructed the world-sheet interfaces which satisfy 
the continuity conditions on the space-time supercharges
in type II  superstring theories in the Green-Schwarz formalism. 
We started with the ansatz  (\ref{susyI}) for the interface gluing type IIB theories, 
which follows from the unfolding  procedure of  the supersymmetric 
 boundary state. The conditions on the 
linearly realized supersymmetry (\ref{QI1}) reduce to those on the 
spinor zero-mode factor. We found two classes of the 
solutions (\ref{MijklFD}) and (\ref{MijklTP}). The conditions 
on the non-linearly realized supersymmetry (\ref{QI2})
reduce to those on the ``S-Matrix'' $S_{AB}$ which determines
the continuity conditions for the oscillator modes.
We in addition required that the continuity conditions induce linear 
transformations of the fields and hence both zero and non-zero
modes transform homogeneously.
We then found a solution  (\ref{SAB}).
The requirement for the  homogeneity of the transformations  
correlated the S-matrix and the spinor zero-mode factor, which
 led to the condition (\ref{aAB}). As a result, we had
two classes of the interfaces (\ref{IFD}) and (\ref{ITP}) in the 
IIB-IIB case.  One class, which is labeled by FD, represents  factorized D-branes.
The other class, which is labeled  by TP,  is regarded as an analog of the topological
interface in two-dimensional CFT, and   found to generate T-dualities.
This result is in accord with the fact that the topological
conformal interface generates symmetries of CFT including
T-dualities \cite{Frohlich:2004ef}. The interfaces gluing IIA theories  
or  IIB and  IIA theory  are similarly obtained.

Having obtained the supersymmetric interfaces, 
we then studied their properties.
First, we observed  that the interface in the TP case  is interpreted as 
a submanifold (``bi-brane'') in the  doubled (transverse) target space
$\bbR^{8} \times \bbR^{8}$, similarly  to the topological 
conformal interface 
\cite{Fuchs:2007fw}. Second, we studied the coupling through the interface
among the NS-NS massless fields, as an example. 
In the TP case,  we found that the coupling 
indeed agrees with the Buscher rules at the linearized level. 
Third, we computed the partition
function with a pair of an interface and its conjugate inserted. 
In a limit, one can read off the spectrum of the modes coupled to the interface.
We found that it is essentially the same as the spectrum for the ordinary D-brane.
In another limit, we also obtained the Casimir energy between the interfaces,
which is regarded as a generalization of the result  for the permeable
interface in two-dimensional CFT \cite{Bachas:2001vj}. 
Finally, we derived the transformations
of the D-branes/supersymmetric boundary states by the interface. 
In the FD case, a D-brane is transformed, when non-vanishing, to the ``left-hand side'' 
of the factorized D-branes represented by the interface.
In the TP case, the transformation is summarized as the multiplication rule 
(\ref{TPMMM})
of the SO(8) matrices which specify the boundary conditions of the fields. 
When the target space is not compactified and the SO(8) matrices 
are those for simple D-branes,  $M^{(p)}$'s in (\ref{Mp}), (\ref{MpIIA}),  
the position moduli in the resultant  D-brane are additive and 
allowed only in the common 
Dirichlet directions.

The supersymmetric interface in the TP case is regarded,  as anticipated, 
as a generator of T-dualities  in type II theories, as well as
an operator acting on the space of the D-branes.
Applications
to the study of non-perturbative aspects and symmetries of superstring theory
would deserve further investigations. 
In this respect, an application 
to solution generating algebras such as 
the U-duality and the Geroch group has been suggested \cite{Bachas}.
Connection to double field theory is also expected from  
the doubling and unfolding procedure in the construction,  
and from  the  interpretation as a ``bi-brane'' in the doubled 
target space.
An interesting   possibility would be that 
the interfaces are realized as solitonic solutions in double field theory, 
similarly to the D-branes/black $p$-branes in supergravity.

Our results in this paper would be extended in several directions.
First, it would be of interest to study whether
 the continuity conditions (\ref{QI1}), (\ref{QI2})
or (\ref{QI1-2}), (\ref{QI2-2}) allow more general solutions,
in particular,  those which connect  the FD and TP cases
as in the permeable conformal interface. 
Second, 
when the target space is compactified, one can expect 
rich structures of the algebras among the interfaces and the D-branes,
as in the fusion of the $c=1$ conformal interfaces \cite{Bachas:2007td}.
This direction should be  explored further. 
Third, in our construction  the conditions on the modes
are of  ``three-term relation'' given by  $S_{AB}$
(suppressing the vector/spinor indices), whereas
those for the supercharges are of ``four-term relation''. The compatibility 
of these two types led to strong constraints,  to leave
the two classes of the  interfaces.
Whether one may consider more general types of the continuity conditions than
(\ref{QI1}), (\ref{QI2}) would be an  issue for future.

The  interfaces constructed in this paper may avoid  the 
difficulty regarding the negative norm states, 
since they preserve two sets of the Virasoro generators.
As  mentioned in \cite{Bachas:2001vj}, whether there could be
exceptions to the argument there deserves further considerations.
This question is closely related to the search for the more
general interfaces discussed above. Finally, at least  
the  supersymmetric interfaces  constructed in this paper
should be realized as conformal interfaces in the RNS formalism of 
superstring theory. In this way, one may study the interface 
in superstring theory (for fixed genus)
in a manifestly covariant manner, though at the cost of some complications
related to space-time supersymmetry and ghosts.

\vspace{3ex}
\begin{center}
  {\large\bf Acknowledgments}
\end{center}

The author would like to thank T. Fujiwara, Y. Imamura, K. Ito, K. Sakai, 
S. Yamaguchi and, especially, S. Hirano for useful discussions. 
This work is supported in part by Grant-in-Aid
for Scientific Research from the Japan Ministry of Education, Culture, 
Sports, Science and Technology. 

\par\bigskip
\appendix

\section{Appendix}

In this appendix, we list the equations which result
from the continuity conditions of the supercharges 
(\ref{QI1}), (\ref{QI2})  in the IIB-IIB case.
First,  acting on the spinor zero-mode factor  (\ref{If0gen}) with the
linearly realized supercharges,  
one finds from (\ref{QI1}) that 
\eqb\label{QI1-2}
   && 0= M_{\adot JK \ddot} \sigmabar^{I}_{\adot a}
      + iR^{1}_{ab} M_{I\bdot K \ddot} \sigmabar^{J}_{\bdot b}
      - M_{IJKL} \sigma^{L}_{a \ddot} 
       +iR^{2}_{ab} M_{IJ\dotc \ddot} \sigmabar^{K}_{\dotc b}
     \comma \qquad  \nn \\
   &&  0= M_{\adot J \dotc L} \sigmabar^{I}_{\adot a}
        + iR^{1}_{ab} M_{I\bdot \dotc L} \sigmabar^{J}_{\bdot b}
        -M_{IJ\dotc \ddot} \sigmabar^{L}_{\ddot a}
        -iR^{2}_{ab} M_{IJKL} \sigma^{K}_{b \dotc}
      \comma \nn \\
    && 0= M_{\adot \bdot K L} \sigmabar^{I}_{\adot a}
        + iR^{1}_{ab} M_{IJK L} \sigma^{J}_{b\bdot}
        - M_{I\bdot K\ddot} \sigmabar^{L}_{\ddot a}
        -iR^{2}_{ab} M_{I\bdot \dotc L} \sigmabar^{K}_{\dotc b}
      \comma \nn \\     
    && 0= M_{IJ K L} \sigma^{I}_{a \adot}
        - iR^{1}_{ab} M_{\adot \bdot K L} \sigmabar^{J}_{\bdot b}
        - M_{\adot J K\ddot} \sigmabar^{L}_{\ddot a}
        -iR^{2}_{ab} M_{\adot J \dotc L} \sigmabar^{K}_{\dotc b}
      \comma \nn \\       
    && 0= M_{I \bdot \dotc L} \sigma^{I}_{a \adot}
        - iR^{1}_{ab} M_{\adot J \dotc L} \sigma^{J}_{b \bdot }
        - M_{\adot \bdot \dotc \ddot} \sigmabar^{L}_{\ddot a}
        -iR^{2}_{ab} M_{\adot \bdot  K L} \sigma^{K}_{b\dotc}
      \comma  \\   
    && 0= M_{I \bdot K \ddot} \sigma^{I}_{a \adot}
        - iR^{1}_{ab} M_{\adot J K \ddot} \sigma^{J}_{b \bdot }
        - M_{\adot \bdot KL} \sigma^{L}_{a\ddot}
        + iR^{2}_{ab} M_{\adot \bdot  \dotc \ddot} \sigmabar^{K}_{\dotc b}
      \comma \nn  \\   
    && 0= M_{I J \dotc \ddot} \sigma^{I}_{a \adot}
        - iR^{1}_{ab} M_{\adot \bdot \dotc  \ddot} \sigmabar^{J}_{\bdot b}
        - M_{\adot J \dotc L} \sigma^{L}_{a\ddot}
        + iR^{2}_{ab} M_{\adot JK \ddot} \sigma^{K}_{b\dotc}
      \comma \nn  \\    
    && 0= M_{\adot \bdot \dotc \ddot} \sigmabar^{I}_{\adot a}
        + iR^{1}_{ab} M_{IJ \dotc  \ddot} \sigma^{J}_{b\bdot}
        - M_{I \bdot \dotc L} \sigma^{L}_{a\ddot}
        + iR^{2}_{ab} M_{I\bdot K \ddot} \sigma^{K}_{b\dotc}
      \period \nn  
\eqe
Next, following the procedure described in the main text, 
a sufficient condition for (\ref{QI2}) to hold turns out to be
\eqb\label{QI2-2}
 \begin{array}{ll}
    0 = \sigma_{a\adot}^{I} \S_{12}^{IJ} - i \sigma_{b\adot}^{J} \S_{12}^{ab}
   \comma 
    & 0 = \sigma_{a\adot}^{I} \S_{12}^{ab} + i \sigma_{b\adot}^{J} \S_{12}^{IJ}
    \comma  \\
     0 = \sigma_{a\adot}^{I} \S_{11}^{IJ} 
         - \sigma_{b\bdot}^{J} R^{1}_{\adot\bdot} \S_{11}^{ab} 
     \comma 
   & 0  = \sigma_{a\adot}^{I} \S_{11}^{ab} 
         - \sigma_{b\bdot}^{J} R^{1}_{\adot\bdot} \S_{11}^{IJ}   
         \comma \\
    0 = \sigma_{a\adot}^{I} \S_{22}^{JI} 
         - \sigma_{b\bdot}^{J} R^{2}_{\adot\bdot} \S_{22}^{ba}       
           \comma 
    & 0 =  \sigma_{a\adot}^{I} \S_{22}^{ba} 
         - \sigma_{b\bdot}^{J} R^{2}_{\adot\bdot} \S_{22}^{JI}       
         \comma  \\
      0 =  \sigma_{a\bdot}^{I} R_{\adot\bdot}^{1} \S_{21}^{JI} 
         + i \sigma_{b\bdot}^{J} R^{2}_{\adot\bdot} \S_{21}^{ba}   
         \comma  \qquad 
     &  0 = \sigma_{a\bdot}^{I} R_{\adot\bdot}^{1} \S_{21}^{ba} 
         - i \sigma_{b\bdot}^{J} R^{2}_{\adot\bdot} \S_{21}^{JI}   
         \period
      \end{array}          
\eqe
The equations in other cases, i.e., in the IIA-IIA and IIB-IIA cases, are obtained
by appropriately changing the chirality and hence index structures.

%
%
\def\thebibliography#1{\list
 {[\arabic{enumi}]}{\settowidth\labelwidth{[#1]}\leftmargin\labelwidth
  \advance\leftmargin\labelsep
  \usecounter{enumi}}
  \def\newblock{\hskip .11em plus .33em minus .07em}
  \sloppy\clubpenalty4000\widowpenalty4000
  \sfcode`\.=1000\relax}
 \let\endthebibliography=\endlist
%
%
\vspace{3ex}
\begin{center}
 {\large\bf References}
\end{center}
\par 

%

%
\end{document}